\documentstyle[aps,prl,twocolumn,epsfig]{revtex} 
\draft
\newcommand{\pathfigs}{.}

\begin{document}

%******************************************************************
%*************** titlepage (abtract) revtex ***********************
%******************************************************************
\tighten
\onecolumn
\twocolumn[\hsize\textwidth\columnwidth\hsize\csname @twocolumnfalse\endcsname
{\protect
\title{Gas-kinetic-based traffic model explaining observed hysteretic
 phase transition}
\author{Dirk Helbing %$^{\ast,\dag}$ 
and Martin Treiber%$^{\ast}$
       }
\address{%$^{\ast}$ 
         II. Institute of 
         Theoretical Physics, University of Stuttgart,
         Pfaffenwaldring 57, 70550 Stuttgart, Germany %,
}
%\date{March 2, 1998}
\maketitle

\begin{abstract}
Recently, hysteretic transitions to `synchronized traffic'
with high values of both density and traffic flow
were observed on German freeways
[B.~S.~Kerner and H.~Rehborn, Phys. Rev. Lett. {\bf 79}, 4030 (1997)].
We propose a macroscopic traffic model based on a gas-kinetic approach
that can explain this phase transition.
The results suggest a general mechanism for the formation
of probably the most common form of congested traffic.
\end{abstract}

} ]
%******************************************************************
% introductionary part
%*******************************************************************
To physicists,
non-equilibrium phase transitions are very fascinating phenomena.
Prominent  examples are pattern-forming transitions in  
hydrodynamic systems driven far from equilibrium, like thermal
convection of a fluid heated from below or
transitions to a state of spatio-temporal chaos \cite{hohen-treiber}.
%A great variety of self-organized phenomena is also found in 
%driven granular media, e.g. the evolution of density waves
%or convection patterns, or the segregation of grains of different sizes
%\cite{pipe}. 
Recently, physicists got interested in the 
spatio-temporal, collective patterns 
of motion formed in social or biological systems 
of so-called `motorized' or `self-driven' particles \cite{mot}. 
A particularly strong physical activity has developed 
in the rapidly growing field of traffic dynamics
\cite{verk,macro,helbing-kinetic,helbing-book,kerner-ramp,kerner-sync,%
kerner-emp2,helbing-emp,ML,treiterer,forthcoming,gaskin,infl,helbing-exp},
not only because of the large potential for industrial applications.

On a macroscopic scale, many aspects of traffic
flow are similar to those of aggregated physical systems.
In particular, if one abstracts from the motion of the single vehicles,
traffic can be modelled as a continuum
compressible fluid \cite{macro,helbing-kinetic}
(see Ref.~\cite{helbing-book} for an overview). 
Existing macroscopic traffic models have been able to explain various 
empirically observed properties of traffic dynamics,
including the transition of %initially homogeneous or
slightly disturbed traffic to traffic jams 
(`local cluster effect') \cite{kerner-ramp}.

Recently, Kerner and Rehborn presented experimental data
indicating a first-order transition to 'synchronized'
traffic (ST) \cite{kerner-sync}. Traffic data from several freeways 
in Germany \cite{kerner-sync,kerner-emp2}
and the Netherlands \cite{helbing-book,helbing-emp} indicate 
that ST is the most common form of congested traffic.
ST typically occurs at on-ramps 
when vehicles are added to already busy `freeways' and
has the following properties:
(i) The dynamics of the average velocities on all lanes is highly correlated
('synchronized').
(ii) ST is characterized by a low average velocity, but, in contrast to
traffic jams, the associated traffic flow is rather high.
(iii) The transition to ST is usually caused by a localized
and short perturbation of traffic flow that 
starts downstream of the 
on-ramp and propagates upstream with a velocity of about $-10$\,km/h.
(iv) As soon as the perturbation passes the
on-ramp, it triggers ST which spreads upstream in the course of time.
(v)  Downstream, ST eventually relaxes to free traffic.
(vi) ST often persists over several hours. 
(vii) The transition from
ST to free traffic occurs at a lower density and higher average 
velocity than the inverse transition {\em (hysteresis effect).}  

Property (i) is related to lane-changing and requires a 
multi-lane model for its description, e.g. \cite{ML}. In order to explain
the other characteristic properties of ST, we 
will propose a macroscopic, effective one-lane model that was derived from
a gas-kinetic level of description and treats all lanes in an overall manner.
The model is also in agreement with other 
empirical findings \cite{kerner-emp2,treiterer} 
like the existence of metastable states, the typical propagation
velocity of upstream jam fronts (between $-10$ and $-20$ kilometers per hour),
and the characteristic outflow $Q_{\rm out}$ from traffic jams
of 1600 up to 2100 vehicles per
hour and lane (depending on the road and weather conditions, but not
on the initial conditions or the surrounding traffic density) 
\cite{forthcoming}.

Our model is based on a kinetic equation for the
phase-space density $\tilde{\rho}(x,v,t)$, 
which corresponds
to the spatial vehicle density $\rho(x,t)$ times the distribution
$P(v;x,t)$ of vehicle velocities $v$ at position $x$ and time $t$
\cite{helbing-kinetic}. (For an introduction to gas-kinetic
traffic models see Ref.~\cite{gaskin}.)
The kinetic equation has some similarities to the gas-kinetic
Boltzmann equation for one-dimensional dense gases 
with the vehicles playing the role of molecules. However, 
there are also some  features specific to traffic. 
Drivers want to accelerate to their respective desired
velocities giving rise to a relaxation term that violates
conservation of momentum and kinetic energy.
Moreover, when approaching a slower car that cannot be immediately overtaken, 
one has to decelerate while the car in front remains unaffected.
This leads to an anisotropic interaction. Finally,
the reaction of the drivers depends on the traffic situation ahead of
them, making the interaction non-local.

The model equations for
the lane-averaged vehicle density
$\rho(x,t) = \int \!dv \,\tilde{\rho}(x,v,t)$
and the average velocity $V(x,t) 
= \rho^{-1} \int \!dv \,v \tilde{\rho}(x,v,t)$ are
\begin{equation}
\label{eqrho} 
 \frac{\partial \rho}{\partial t} + 
 \frac{\partial (\rho V)}{\partial x} 
 = \frac{Q_{\rm rmp}}{nL} \, ,
\end{equation}
\begin{eqnarray}
\label{eqV}
  \left( \frac{\partial}{\partial t} 
 + V \frac{\partial}{\partial x} \right) V
   &=& 
 - \frac{1}{\rho} \frac{\partial (\rho\theta)}{\partial x}
       + \frac{V_0-V}{\tau} \nonumber \\
       && -  \frac{V_0 A(\rho) (\rho_{\rm a} T V)^2}
         {\tau A(\rho_{\rm max}) (1-\rho_{\rm a}/\rho_{\rm max})^2}  
  B(\delta_V) , \!\!
\end{eqnarray}
where we use the notation $f_{\rm a}(x,t) \equiv f(x_{\rm a},t)$ 
with $f\in\{\rho,V,\theta\}$ and an advanced `interaction point' $x_{\rm a}$
specified later.
Without on- or off-ramps, the density equation (\ref{eqrho}) is just
a one-dimensional continuity equation reflecting the 
conservation of the number of vehicles. 
Thus, the temporal
change $\partial \rho/\partial t$ of the vehicle density is just given
by the negative gradient $-\partial Q/\partial x$ of the
lane-averaged traffic flow $Q = \rho V$.  
Along on-ramps (or off-ramps), the source term $Q_{\rm rmp}/(nL)$
is given by the actually observed inflow $Q_{\rm rmp}>0$ 
from (or outflow $Q_{\rm rmp}<0$ 
to) the ramp, divided by
the merging length $L$ and by the number $n$ of lanes.
The inflow has an upper limit %$Q_{\rm in}^{\rm max}$ 
that depends on the
downstream flow on the main road \cite{infl}.

The velocity equation (\ref{eqV}) contains the velocity
variance $\theta(x,t)= 
\rho^{-1} \int \!dv \,[v-V(x,t)]^2 \tilde{\rho}(x,v,t)$.
Instead of deriving a dynamic equation for $\theta$ from the kinetic
equations, 
we use the constitutive relation $\theta = A(\rho)V^2$ with
\begin{equation}
 A(\rho) =  A_0 + \Delta A \left[ 1 + \tanh 
      \left(\frac{\rho -\rho_c}{\Delta\rho} \right) \right] \, ,
\end{equation}
where $A_0=0.008$, 
$\Delta A = 0.015$,
$\rho_{\rm c} = 0.28\rho_{\rm max}$,
and $\Delta\rho = 0.1\rho_{\rm max}$ \cite{forthcoming}.
These coefficients can be obtained from single-vehicle
data. Unfortunately, no such data were available for the motorway
considered in \cite{kerner-sync}, but
similar values were obtained for another  motorway
\cite{helbing-exp}. 
\par
The first term on the rhs of Eq.~(\ref{eqV}) 
is the gradient of the `traffic pressure' $\rho\theta$.
It describes the kinematic dispersion of
the macroscopic velocity in inhomogeneous traffic as a consequence
of the finite velocity variance. For example, the macroscopic velocity
in front of a small  vehicle cluster will increase {\it even if no
individual vehicle accelerates}, because the faster
cars will leave the cluster behind. 
The second term denotes the acceleration towards
the (traffic-independent) average desired velocity $V_0$ 
of the drivers with a relaxation time $\tau\in$ [10\,s,50\,s].
Individual variations of the desired velocity are accounted for by
a finite velocity variance.
The third term of the rhs of Eq.~(\ref{eqV}) models braking
in response to the traffic situation
at the advanced `interaction point' 
$x_{\rm a} = x + \gamma( 1/\rho_{\rm max} + T V)$.
In dense traffic, where most drivers maintain the
safety distance $T V$, this point is about $\gamma$ vehicle positions
in front of the actual vehicle position $x$.
The average safe time headway $T$ is of the order of one second.
For the `anticipation factor' $\gamma$, we assume values between one and
two. The braking deceleration increases coulomb-like with decreasing 
gap $(1/\rho_{\rm a} - 1/\rho_{\rm max})$ to the car in front
($1/\rho_{\rm a}$ being the average distance between successive vehicle positions,
$1/\rho_{\rm max}$ the average vehicle length, and
$\rho_{\rm max}$ the maximum density).
In homogeneous dense traffic, the acceleration and braking terms 
compensate for each other at about the safe distance. 
In general, the deceleration tendency depends also on the
velocity difference to the traffic at the interaction point.
A gas-kinetic derivation leads 
to the `Boltzmann factor' \cite{forthcoming}
\begin{equation}  
\label{B}
B(\delta_V) =   2 \left[ 
    \delta_V \frac{\mbox{e}^{-\delta_V^2/2}}{\sqrt{2\pi}}
           + (1+\delta_V^2) 
    \int_{-\infty}^{\delta_V} dy \, \frac{\mbox{e}^{-y^2/2}}{\sqrt{2\pi}}
                 \right],
\end{equation}
where
%
%\begin{equation}
%\label{deltaV}
$\delta_V = %\frac{V(x,t) - V(x_{\rm a},t)}
%            {\sqrt{\theta(x,t)+\theta(x_{\rm a},t)}} \equiv
(V-V_{\rm a})/\sqrt{\theta+\theta_{\rm a}}$
%\end{equation}
%
is the dimensionless velocity difference between the actual location $x$ 
and the interaction point $x_{\rm a}$. In homogeneous traffic, we have $B(0)=1$.
If the preceding cars are much slower (i.e. $\delta_V \gg 0$), it follows
$B(\delta_V) = 2 \delta_V^2$. In the opposite case (i.e. $\delta_V \ll 0$), 
we have $B(\delta_V) \approx 0$. That is, since the distance is increasing, 
then, the vehicle will not brake, 
even if its headway is smaller than the safe distance.
\par
In contrast to previous approaches, the above macroscopic traffic model
explicitly contains an anisotropic, non-local interaction term
$B(\delta_V)$. This is
not only essential for a realistic treatment of
situations with large gradients of $\rho(x,t)$ or $V(x,t)$,
but also for an efficient and robust numerical integration.
Moreover, the prefactor of $B$ has now been obtained from the plausible
assumption that, at high densities, the time headway between successive 
vehicles is $T$. Finally, all model parameters are meaningful, measurable,
and have the correct order of magnitude.

Our simulations have been carried out 
%Let us now turn to the discussion of simulation results, which 
with an explicit finite-difference 
integration scheme and the following parameter values:
$V_0=128$\,km/h,  
$\rho_{\rm max}=160$\,vehicles/km, 
$T=1.6$\,s,
$\tau=31$\,s, 
and $\gamma=1.0$.
The response of equilibrium traffic to
localized disturbances 
is similar to the Kerner-Konh{\"a}user model
\cite{kerner-ramp}. For densities 
$\rho < \rho_{\rm c1}$ %=17$ \,vehicles/h  
and $\rho > \rho_{\rm c4}$, %=57.5$\,vehicles/h,  
homogeneous traffic is
stable, and for a range %$19\,\mbox{vehicles/h} =
$\rho_{\rm c2} < \rho < \rho_{\rm c3}$ %=51.5\,\mbox{vehicles/h}$
of intermediate densities, it is 
linearly unstable, giving rise to cascades of traffic jams 
(`stop-and-go traffic'). For the two density regimes 
$\rho_{\rm c1} \le \rho \le \rho_{\rm c2}$ and 
$\rho_{\rm c3} \le \rho \le \rho_{\rm c4}$ between the stable and
the linearly unstable regions, it is metastable, i.e., it behaves
nonlinearly unstable with respect to perturbations exceeding a 
certain critical amplitude, but otherwise stable. For 
the self-organized density $\rho_{\rm jam}$ inside traffic jams 
we find a typical value $\rho_{\rm jam} > \rho_{\rm c4}$ \cite{forthcoming}.
%, here $\rho_{\rm jam} = 66$\,vehicles/km.

Now, we will discuss synchronized flow. 
Figure~1 shows the simulation of freeway traffic 
near an on-ramp during a `rush-hour', 
where we assumed that the flow downstream of the on-ramp
almost reaches the maximum equilibrium flow (`capacity limit')
$Q_{\rm max}$.
The upstream boundary condition at position $x_0 = -6$\,km
was specified in accordance with the
equilibrium flow-density relation 
for free traffic (dotted lines in Fig.~3, before the maximum of the curve)
with flows according to Fig.~1(c).
We started with a high main flow that is monotonically decreasing
in the course of time. At $x=0$\,km,
an on-ramp with merging length $L = 300$\,m 
injects an additional time-dependent inflow $Q_{\rm rmp}$
into the freeway. This on-ramp flow was assumed to have a short and tiny peak
at $t=10$\,min [Fig.~1(c)].
As a result, a wave of denser traffic propagated downstream,
thereby gaining a larger amplitude,
and eventually propagated upstream 
again with about $-11$\,km/h. 
Once the perturbation reached the ramp, 
dense traffic (of about 48 vehicles/km) with
relatively high flows (1600 vehicles/h) corresponding to $V=$\,33 km/h
built up in the upstream direction. Although the flow from the main road 
was gradually decreased for $t>$ 30 min, 
it took more than 100 additional minutes, until the
region of congested traffic vanished.

%******************************************************************
%  Vergleich mit Kerners PRL
%*******************************************************************

All these features agree with
the experimental observations of ST described in
Ref.~\cite{kerner-sync}.
There, a peak on the on-ramp flow was observed at about 7:15\,am. 
The transition to ST
was first detected at 7:16\,am as a short dip of the velocity
700m downstream from the on-ramp 
(detector D3 in \cite{kerner-sync}).
At about 7:22\,am, the front reached a detector (D2) 200m upstream
of the ramp (corresponding to a mean propagation speed of $-11$\,km/h), 
and propagated slower to the next detector D1 (700m upstream).
While the perturbation at detector D3 lasted only a few minutes, 
it was followed 
by nearly 2 hours of congested traffic ($V \approx 30 $\, km/h,
$Q \approx $\,1500 vehicles/h)
at the detectors D2 and D1.

%We have compared the empirical findings and simulations in more detail.
It turned out that, apart from fluctuations, 
the simulated velocities and flows obtained
at the detector positions $x=-0.7$\,km (D1), $x=-0.2$\,km (D2),
$x=0.7$\,km (D3), and $x=1.5$\,km (D4) (cf. Fig.~2)
are in almost quantitative agreement with
all features of ST as displayed in
Figs.~2(c), 2(b), 2(a), and 2(d)
of Ref.~\cite{kerner-sync}.
In particular, the model reproduces the drop of the velocity
to about 30 km/h for up to two hours, while
the flow is reduced by only 20\%. 
Moreover, after the transition to free flow, 
the velocity is higher and the flow
is lower than immediately before the transition to synchronized flow, 
both in the measurements and the simulation.
Finally, in Fig.~3 we depict
the relaxation to free traffic downstream of the ramp
by flow-density diagrams [see also Fig.~1(b)].
The results agree well with the empirical traffic data presented
in Fig.~3(c) of  Ref.~\cite{kerner-sync}.

%******************************************************************
%  Diskussion: Vorgeschlagene Erklaerung des ST und Ausblick
%*******************************************************************

Our results suggest the following interpretation of 
the phase transition to
%the mechanism behind the formation of 
ST.
Initially, the homogeneous flow $Q_{\rm main}$ upstream of an on-ramp 
is stable, while the higher downstream flow
$Q_{\rm down} = Q_{\rm main} + Q_{\rm rmp}/n$ is metastable ($n=$
number of lanes). A  perturbation of  the ramp flow $Q_{\rm rmp}$ 
triggers a stop-and-go wave, which travels downstream as long as it is 
small and upstream as it becomes larger, as is 
known from 'localized clusters' \cite{kerner-ramp}. 
Now, assume the downstream front of the cluster would
pass the on-ramp. Then, since $Q_{\rm main}$ [Fig.~1(c)] is lower
than the characteristic outflow $Q_{\rm out}$ 
from a jam (being of the order of 2000 vehicles/km),
the cluster would eventually vanish. 
However, during its lifetime, the cluster would continue to emit
the flow $Q_{\rm out}$, leading downstream of the ramp to a 
flow $Q_{\rm out}+Q_{\rm rmp}/n > Q_{\rm max}$.
As a consequence, 
as soon as the perturbation reaches the on-ramp, 
it induces congested traffic with a {\em standing}
downstream front just at the end of the ramp.
With an observed outflow 
$\tilde{Q}_{\rm out} \lesssim Q_{\rm out}$ 
from ST \cite{Notice},
the average flow upstream is given by
\begin{equation}
\label{qsync}
Q_{\rm sync} = \tilde{Q}_{\rm out} - Q_{\rm rmp}/n \, .
\end{equation}
Now, consider the density $\rho_{\rm sync}$ defined by
$Q_{\rm sync} = Q_{\rm e}(\rho_{\rm sync})$ in the congested part of 
the equilibrium flow-density relation $Q_{\rm e}(\rho)$
(dotted lines in Fig.~3, behind the maximum of the curves). 
If homogeneous traffic is (me\-\mbox{ta-)}stable 
at $\rho_{\rm sync}$, the on-ramp 
induces ST, otherwise it induces dynamically changing states. 
The restriction $Q_{\rm rmp} %\le Q_{\rm in}^{\rm max}
\le \tilde{Q}_{\rm out}/2$ \cite{infl} (corresponding to every second 
vehicle on the right freeway lane stemming from the on-ramp) implies
$Q_{\rm sync} \ge (1 - \frac{1}{2n})\tilde{Q}_{\rm out}$ and
$\rho_{\rm sync} < \rho_{\rm jam}$, so that
synchronized flow is significantly higher than the flow
inside traffic jams. 
%******************************************************************
% Schluss
%*******************************************************************

We have proposed a macroscopic traffic model based on a gas-kinetic
level of description that allows to describe the empirically observed
features of traffic flows. This Letter focussed on the simulation and
interpretation of ST, which is probably the most common
form of congested traffic. We have triggered ST
by a small peak in the inflow from an on-ramp, when the downstream flow 
was close to freeway capacity. 
Synchronized traffic
eventually resolved in downstream direction, but spread in upstream
direction. It persisted for more than one hour, although the
main flow was steadily reduced. We also performed simulations 
without peaks, leaving everything else unchanged. In these cases, 
we obtained free traffic flow. This confirms that 
the proposed model can describe the hysteretic and bistable properties of
real traffic. Our interpretation of ST 
underlines the crucial role of the characteristic outflow $Q_{\rm out}$ 
from congested traffic for traffic dynamics.
The simple criterion $Q_{\rm out} + Q_{\rm rmp}/n > Q_{\rm max}$ 
for the formation of ST can be useful for
determining bottlenecks of the existing road infrastructure
as well as for planning efficient freeway networks.

%******************************************************************
%  Acknowledgements
%*******************************************************************

The authors want to thank for financial support by the BMBF (research
project SANDY, grant No.~13N7092) and by the DFG (Heisenberg scholarship
He 2789/1-1).

%******************************************************************

%******************************* Fig. 1 ************************************
\unitlength=0.5mm
\begin{figure}
\label{fig1}
\begin{center}
   \includegraphics[width=160\unitlength]
   {\pathfigs/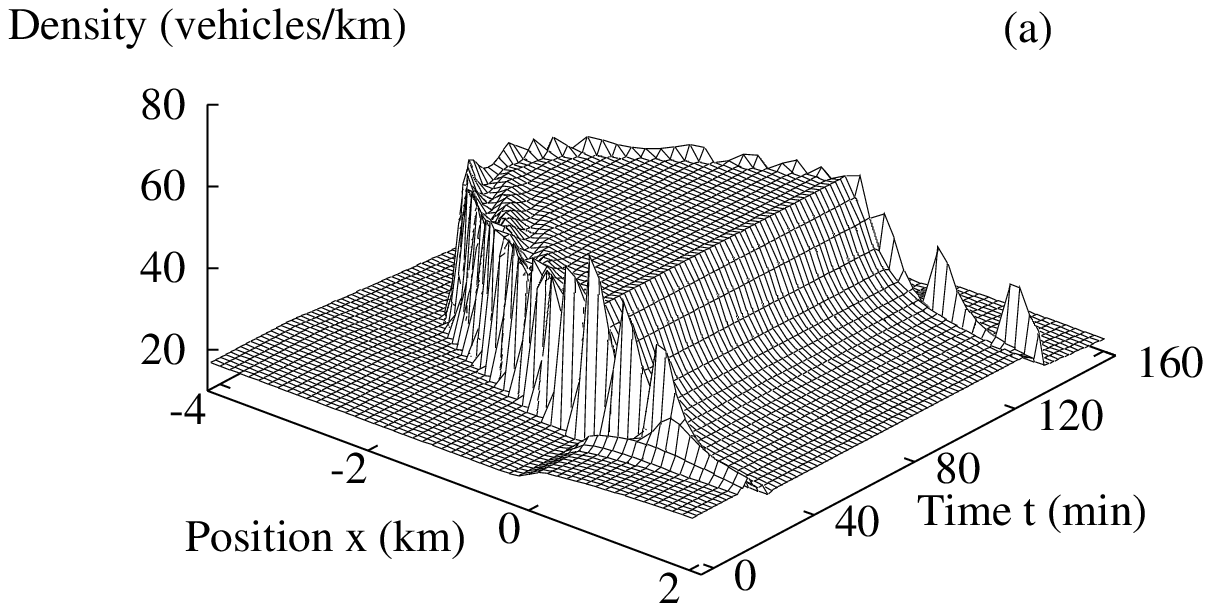} \\[-20\unitlength]
   \includegraphics[width=95\unitlength]{\pathfigs/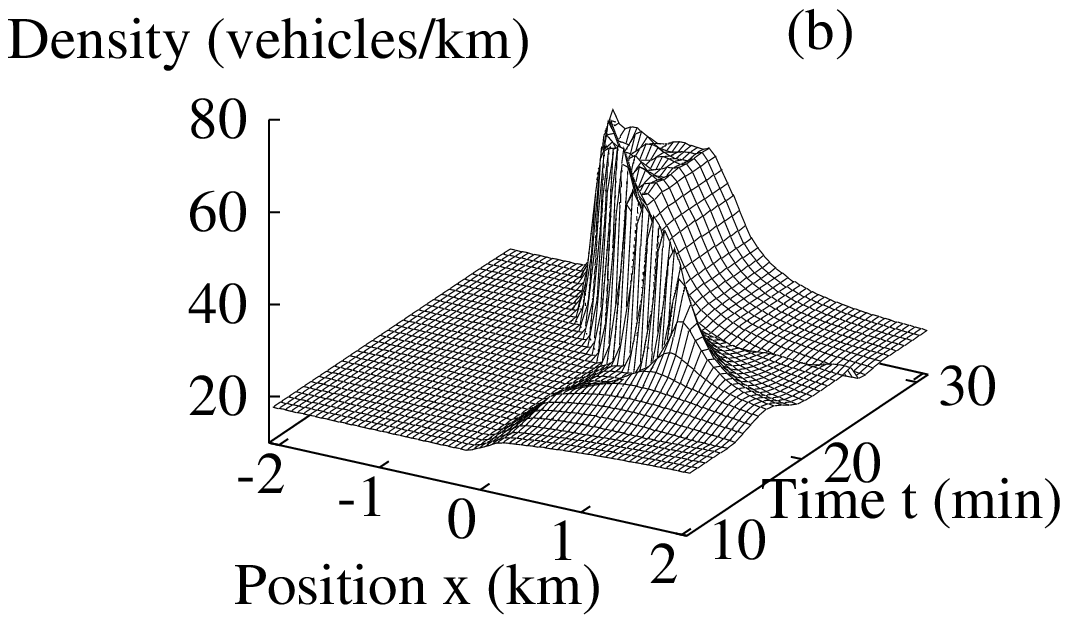} 
   \hspace{-10\unitlength}\vspace*{-5\unitlength}
   \includegraphics[width=65\unitlength]{\pathfigs/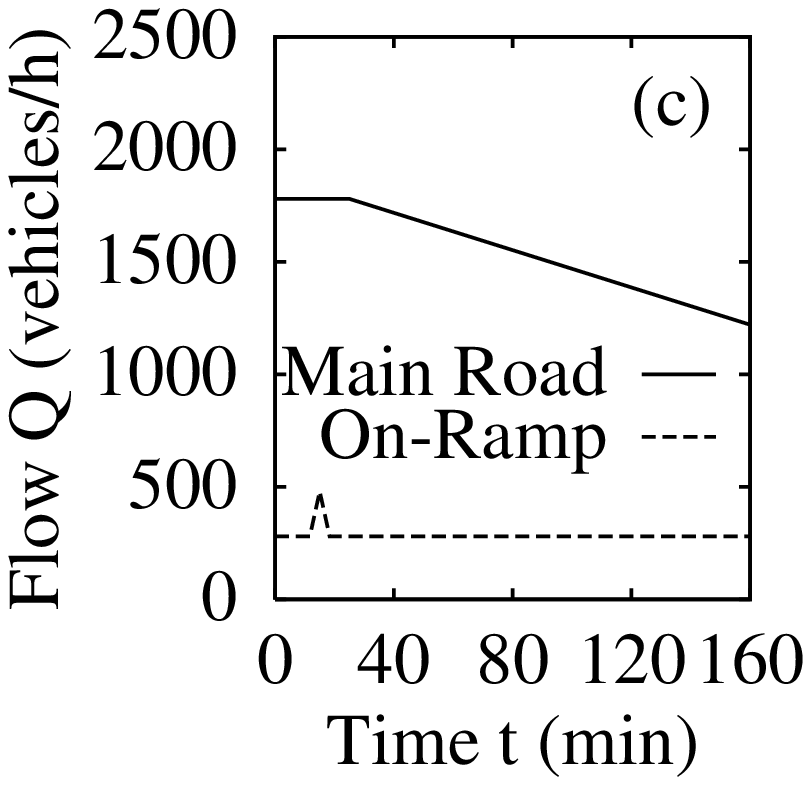} 

\end{center}
\caption{Spatio-temporal evolution of the lane-averaged density 
after a small peak of inflow from the on-ramp. The on-ramp merges with
the main road at $x=0$\,km with a merging length of 300m.
Traffic flows from left to right.
In (a), the parabolically shaped region of high density
corresponds to ST.
Plot (b) shows the formation of this state in more detail.
The time-dependent inflows $Q_{\rm main}$ at the upstream boundary 
and $Q_{\rm rmp}/n$ at the on-ramp are displayed in (c).} 
\end{figure}

%******************************** Fig. 2 ******************************

%\newpage
\unitlength0.6mm
\begin{figure}
\label{fig2}

\vspace{0\unitlength}

\begin{center}

\hspace*{-10\unitlength}\includegraphics[width=150\unitlength]
 {\pathfigs/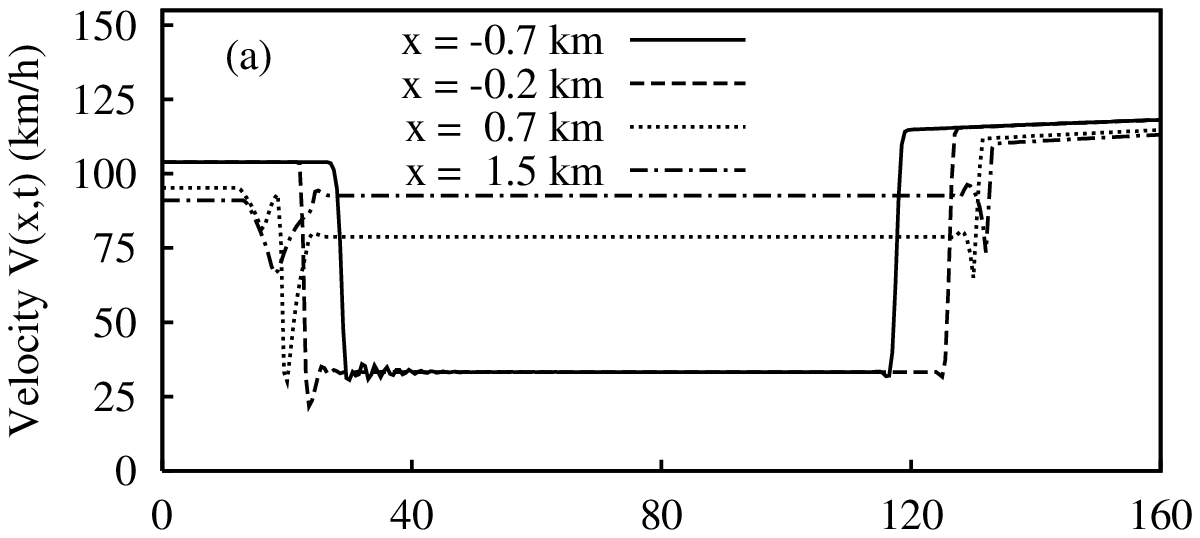} \\[-5\unitlength]
\hspace*{-10\unitlength}\includegraphics[width=150\unitlength]
 {\pathfigs/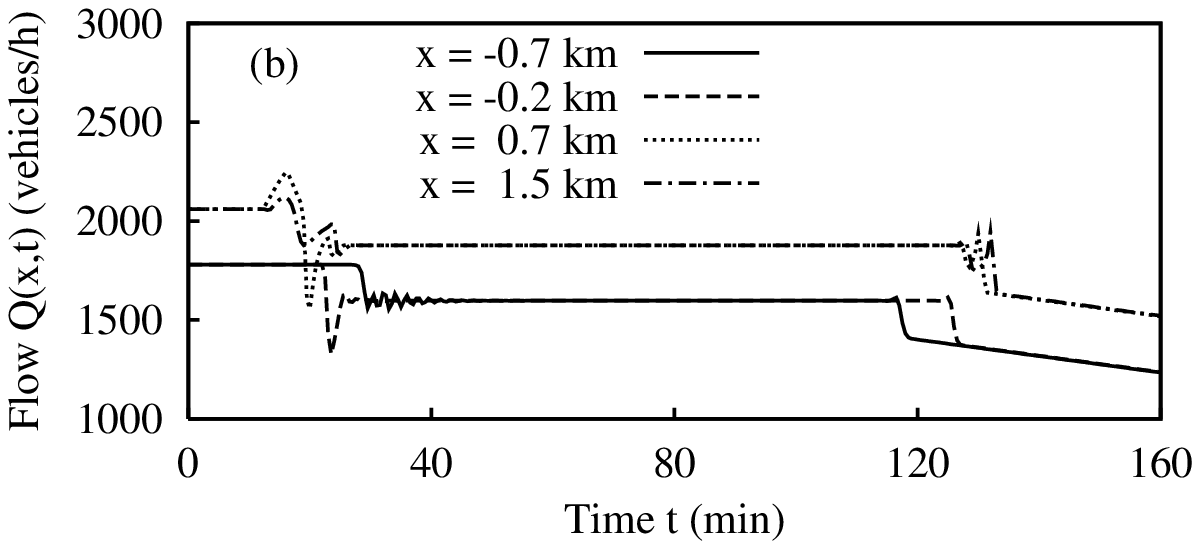}

\end{center}

\caption[]{Temporal evolution of (a) the average velocity and (b) the
traffic flow per lane at four cross sections of the freeway near the on-ramp.
In front of the on-ramp ($x<0$), 
ST exists for a certain time interval. 
Downstream ($x>0$), the traffic
situation recovers towards a freely flowing state. The simulated
overshooting at the beginning of the breakdown of average velocity
is in agreement with empirical observations (cf. Fig.~1(b) in
Ref.~\protect\cite{kerner-sync}).} 

\end{figure}

%************************** Fig. 3 ********************************

%\newpage

\unitlength0.55mm
\begin{figure}
\label{fig3}

\vspace{0\unitlength}

\begin{center}
\hspace*{-10\unitlength}\includegraphics[width=58\unitlength]
 {\pathfigs/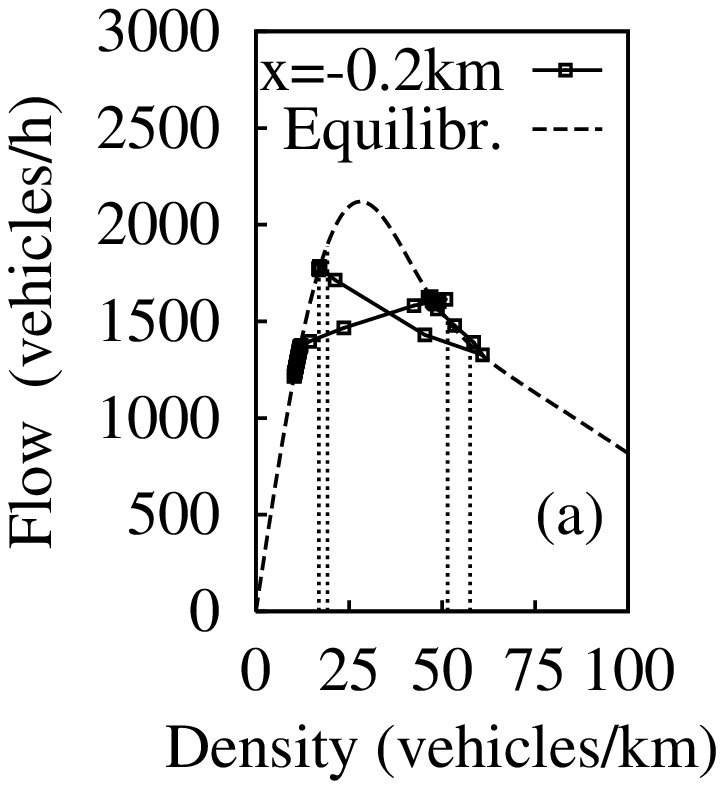}  \hspace{-10\unitlength}
   \includegraphics[width=58\unitlength]{\pathfigs/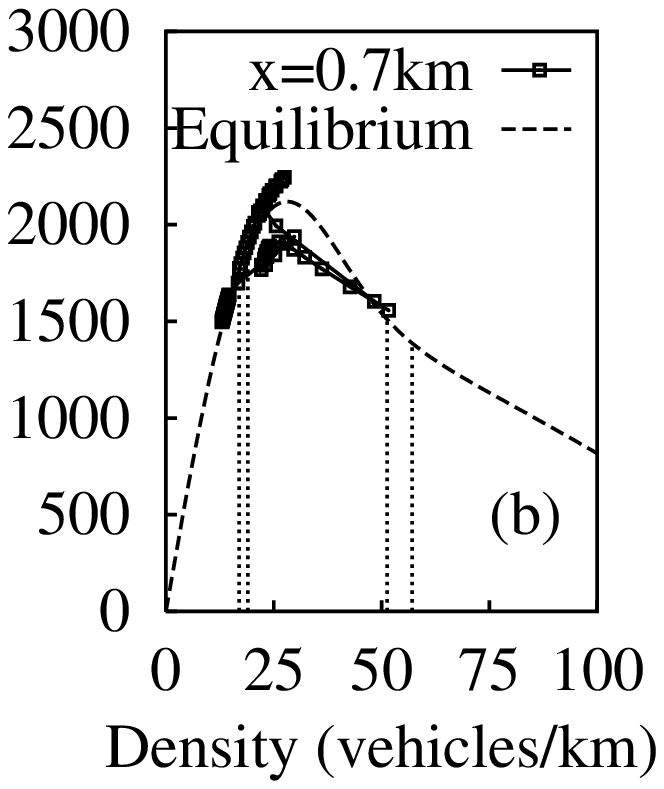}  
 \hspace{-10\unitlength}
   \includegraphics[width=58\unitlength]{\pathfigs/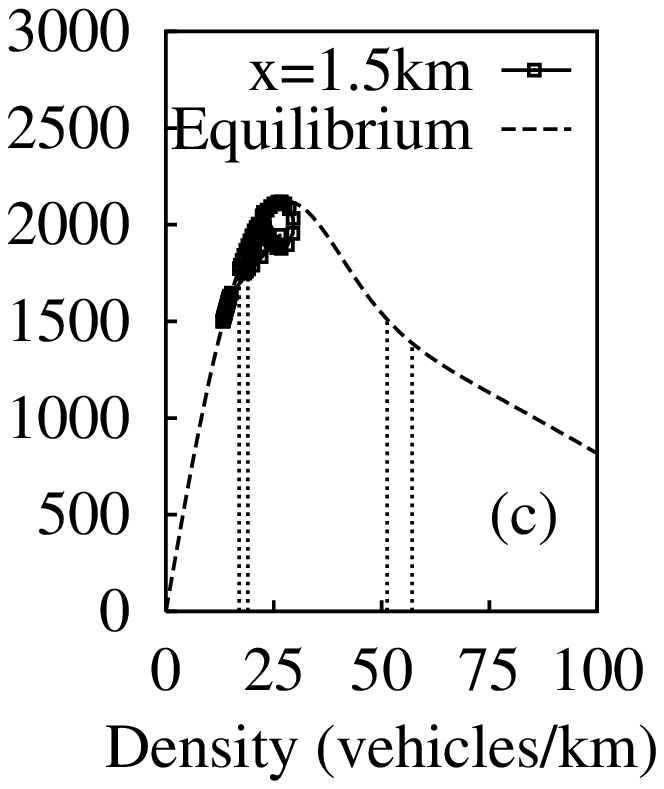}  
 \hspace{-10\unitlength}
\end{center}

\caption{
Traffic dynamics in the flow-density plane (a)
0.2 km upstream of the on-ramp and (b), (c) at two downstream cross sections. 
The solid lines with the symbols ($\Box$) correspond to the simulation 
results of Fig.~1. All the trajectories start 
at $\rho=17$\,vehicles/km and $Q=1770$\,vehicles/h.
The dashed line represents the equilibrium relation $Q_{\rm e}(\rho)$
of the model. The vertical dotted lines indicate the stability limits
$\rho_{\rm c1}$, $\rho_{\rm c2}$, $\rho_{\rm c3}$, and $\rho_{\rm c4}$
(determined numerically).
}
\end{figure}

\end{document}